\documentstyle[aps,preprint,epsf]{revtex}
\newcommand{\be}{\begin{equation}}
\newcommand{\ee}{\end{equation}}
\begin{document}
\draft
\title{Monte Carlo Simulation of \\
Sinusoidally Modulated Superlattice Growth \\} 
\author{H. Jeong$^1$, B. Kahng$^{1,2,*}$, S. Lee$^{1}$, C.Y. Kwak$^3$, 
A.-L. Barab\'asi$^1$, and J.K. Furdyna$^1$ \\}
\address{
$^1$ Department of Physics, University of Notre Dame, Notre Dame, IN 46556 \\
$^2$ School of Physics and Center for Theoretical Physics, Seoul National 
University, Seoul 151-742, Korea \\
$^3$ Department of Physics and Center for Advanced Materials 
and Devices, Konkuk University, Seoul 143-701, Korea \\}
\maketitle
\renewcommand\baselinestretch{1.35}
\thispagestyle{empty}
\begin{abstract}
The fabrication of ZnSe/ZnTe superlattices grown by the process 
of rotating the substrate in the presence of an inhomogeneous flux 
distribution instead of successively closing and opening 
of source shutters is studied via Monte Carlo simulations. 
It is found that the concentration of each compound 
is sinusoidally modulated along the growth direction, 
caused by the uneven arrival of Se and Te atoms at a given 
point of the sample, and by the variation of the Te/Se ratio at 
that point due to the rotation of the substrate. In this way we 
obtain a ZnSe$_{1-x}$Te$_x$ alloy in which 
the composition $x$ varies sinusoidally along the growth direction. 
The period of the modulation is directly controlled by the rate of  
the substrate rotation. The amplitude of the 
compositional modulation is monotonous for 
small angular velocities of the substrate rotation, 
but is itself modulated for large angular velocities. 
The average amplitude of the modulation pattern decreases 
as the angular velocity of substrate rotation increases 
and the measurement position approaches the center of rotation. 
The simulation results are in good agreement with previously 
published experimental measurements on superlattices fabricated 
in this manner. 
\end{abstract}


Superlattices based on II-VI semiconductors are receiving 
continued attention as a result of the wide range of new physical 
phenomena observed in these systems. 
In particular, several superlattice such as 
ZnSe/ZnTe \cite{prb1,prb2}, ZnSe/MgS \cite{mgs}, and 
ZnSSe/ZnSe \cite{znsse} 
have been found interesting because of their potential for 
applications in optical devices at short wavelengths. 
Such superlattices are typically fabricated 
using molecular beam epitaxy (MBE), achieving a periodic structure 
by repeated openings and closings of the source 
shutters. This method automatically results in the 
formation of abrupt interfaces, so that the wells and barriers 
forming the superlattice are shaped firmly a square pattern. \\ 

Growing superlattices 
with profiles deviating from a square composition profile is desirable 
for both fundamental research and potential applications. 
Motivated by this interest, recently a new model of fabricating ZnSe/ZnTe 
superlattices has been introduced, using the rotation of the 
substrate as the modulating mechanism instead of shutter opening and 
closing \cite{prb1}, resulting in sinusoidally modulated superlattices 
(SMSLs)\cite{prb1}. 
The compositional profile of this new structure has been 
investigated using X-ray diffraction studies, 
which systematically exhibit a main alloy peak arising from the  
average ZnSe$_{1-x}$Te$_x$ lattice constant of the alloy, with only one 
satellite on each side of the main peak, the latter  
implying that the chemical composition of the alloy $x$ 
varies sinusoidally along the growth direction \cite{xray}. \\

The formation of such SMSLs results from the rotation of the substrate 
in the presence of a flux inhomogeneity. 
Such inhomogeneity can arise, e.g., from different distances 
of the Zn, Se, and Te sources 
from a given position on the substrate at any one instant. 
The fact that the period of the sinusoidal modulation is 
proportional to the period of the rotational motion 
of the substrate is unambiguously indicated by X-ray measurements, 
the separation of the superlattice satellites from the main peak 
indicating that the period grows with the rotation rate. 
It should be noted that the satellite peak 
is as narrow as the main alloy peak, indicating that the SMSL 
structures are of very good quality \cite{prb1}. 
The band structure of the SMSL is interesting 
as well, different from that of $standard$ superlattice with 
abrupt interfaces \cite{prb3}. 
Accordingly, SMSLs are promising for the observation of new optical 
effects, especially those involving transitions that are forbidden 
by standard selection rules, and possibly for new optical devices. 
Even though the physical properties of the SMSL structures 
are of considerable interest, little has been done so far to firmly 
establish the process of their formation in any quantitative sense.
In this paper, we perform Monte Carlo (MC) simulations reproducing 
the fabrication process of the SMSLs aiming to understand the basic 
mechanism contributing to the formation of those structures, 
helping us to define their structural properties. \\
 
$Experiment$: A ZnSe$_x$Te$_{1-x}$ SMSL was fabricated 
using a Riber 32 R \& D MBE machine, using 
elemental solid sources, Zn, Se, and Te. 
Each source cell is located at an approximate distance of 12 cm 
from the center of the mounting block (``the stage"). 
The Se and Te sources are located 
on each side of the Zn source, and the flux rates from the 
Se and Te are adjusted to make their respective amounts arriving 
at the center of the stage the same. 
The relative positions of the sources are shown schematically in Fig.1. 
The superlattices were grown on (100) GaAs substrates mounted with 
indium on the molybdenum mounting block, having a 
diameter of 5.6 cm. 
The growth of the SMSL was monitored via reflection 
high electron diffraction (RHEED), whose streaky appearance 
indicated good layer-by-layer growth throughout the SMSL growth 
process.\\ 

The configuration of the sources shown in Fig.1 produces a 
nonuniform distribution of flux arriving across the substrate 
(Se-rich on one side, Te-rich on the other). 
The consequences of this inhomogeneity manifest themselves clearly 
when the GaAs substrate has the shape of a long strip (see Fig.1), 
mounted so that one end of the substrate is much 
close to the Se source, and the other to the Te source. 
A strip sample grown in this way without rotation indeed shows a continuous 
gradient of the concentration along the strip, measured via
the X-ray diffraction at various positions along the strip. 
The maximum difference in the Te and Se concentrations 
at the two edges of the strip was measured to be approximately 30\%,
while the concentrations of the two anions at the center of 
the strip substrate were equal, resulting in a homogeneous 
ZnSe$_{0.5}$Te$_{0.5}$ alloy. \\   

$Simulations$: In performing the MC simulations we account 
for the geometry of the sources and the substrate in the 
chamber as follows: We choose an anisotropic cell array of 
$L_x \times L_y =280 \times 20$, corresponding to 
5.6$\times$0.4 cm substrate size, where either ZnSe or ZnTe is 
deposited on each cell. The size of a unit cell is $0.2\times 0.2$mm, 
and $a=0.2$mm is taken as the ``lattice constant" of the cell.
Following this scale, we assume that the sources of 
Zn, Se, and Te are positioned on a circle with the radius of 600$a$ 
(corresponding to 12cm) from the center of the substrate, 
so that the ratio between the diameter of the substrate 
and the distance to the sources is invariant to the experimental 
values. We use a three-dimensional reference frame whose 
origin is taken at the center of the substrate, with the 
growth direction taken as the $z$ direction. $x$-axis ($y$-axis) 
is taken along the longitudinal (transverse) direction of the 
strip substrate (see Fig.~1). The substrate is then on the $z=0$ 
plane, and the sources are located at $S_i=(x_i, y_i, z_i)$, 
where the index $i=0$,~1 and 2 stands for Zn, Se, and Te, 
respectively. Thus, for example, the Zn source is located at 
$S_0=(-r_0\sin 11^{\circ}, 0, r_0\cos 11^{\circ})$, with 
$r_0=600a$.   
In the actual experiment the positions of the sources are 
fixed in the $(x,z)$-plane, while the substrate rotates with 
the angular velocity $\omega$. In our simulations we fix  
the substrate and equivalently rotate the sources in 
the reverse direction with the same frequency $\omega$.
The epilayer height $h(x,y)$ at any time is defined as the height 
of the top of the epilayer at the position $(x,y)$ on the substrate. 
The flux from the source $i$ at 
the position $P=(x,y,h)$ on the surface is given by 
\begin{equation}
F_i ={f_i \over 4 \pi}{{\cos \theta_i }\over {r_i^2}}=
{f_i \over 4 \pi}{{|z_i-h|}\over {r_i^3}}, 
\end{equation}
where $f_i$ is the emission rate of the source $i$,  
$r_i$ is the distance between the source and position $P$, 
and $\theta_i$ is the polar angle between the vertical line 
of the source $i$ and the line connecting the source $i$ 
and the point $P$. The emission rates $f_i$ of the Se and 
Te sources are adjusted so as to make the Se and Te fluxes 
at the center of the substrate identical. In simulations, 
we choose $f_1=1.0$. Accordingly, $f_2=\cos(33^o)/\cos(11^o)$, 
leading to the total emission rate $f_1+f_2\approx 1.854$.\\ 
  
The MC simulations are performed using discretized time, 
where a unit time step is taken to be 
$\Delta t=1/[(f_1+f_2)L_xL_y]$. Then the time for one monolayer 
to grow is $T_m=1/(f_1+f_2)$. 
At each time step $\Delta t$, the following calculations 
are carried out. 
First, we choose a single cell among the $280\times 20$ cells 
at random, and a single molecular species of either ZnSe or ZnTe is 
deposited on that cell. Second, the selection of either ZnSe or ZnTe 
for the deposition is determined according to the probability 
proportional to the ratio of the two fluxes arriving at the selected 
cell from the Se and Te sources. Third, the selected molecular 
species is allowed to diffuse into the lowest-lying cell within the 
nearest-neighboring cells. If there are more than one lowest cells, 
then one of those cells is randomly selected with equal probability.
The substrate is then allowed to rotate with an angle 
$\Delta \theta=\omega \Delta t$, after which the above 
three MC simulation steps are repeated. \\ 
 
The first three MC simulation steps incorporated in our model 
are based on the following physical reasoning: First, since each cell has 
a linear size of 0.2 mm, which is enormous on the atomic scale, 
in the experiment a large number of compound molecules can be deposited 
within a single cell. However, it is not feasible 
to deal with such large numbers of molecules in the simulations. 
Thus in the MC simulations we deal with a single representative 
of the large number of compound molecules in a unit cell. Consequently, 
molecular species in the simulations represent a coarse-grain 
particle, comprised of a large number of molecules on the atomic 
scale\cite{wolf}. 
However, in order to see the modulation pattern 
along the $z$ direction, we use a disproportionate short unit 
length scale for the $z$ direction. Specifically, the height 
of the unit cell is taken 
as small as $5 \times 10^{-6}$ times the lateral size 
of the cell, so that the upward growth of the superlattice is simulated 
in terms of 1024 layers. This corresponds to a height of $\sim 1\mu m$, 
the thickness of superlattices fabricated in a typical experiment. \\
 
The second step is based on the assumption that the Zn element is 
supplied in sufficient amount to react with both Se and Te, and the 
reaction rates of the Zn-Se and Zn-Te 
are identical, even though in reality the reaction rate of Zn-Se 
is known to be slightly higher than that of the Zn-Te. 
When the reaction rates are assumed to be identical, a coarse-grained 
species can be selected as the ``majority alloys" within 
a given cell, proportional to the ratio of the incoming fluxes. \\
 
For the third step, we consider the following situation: when Zn, Se, 
and Te atoms arrive at the epilayer surface, they have sufficient 
energy to diffuse into the lowest-lying sites within the diffusion 
length, followed by nucleation, and they then remain at these most 
stable sites. In the simulation, since we deal with the coarse-grain 
particles instead of individual adatoms, their collective behavior 
is very complicated. In reality, the collective motion is not contained 
within the cell, but mass flow occurs across the boundary between neighboring 
cells. Accordingly, to mimic this mass flow, we allow incoming 
coarse-grain particles to diffuse to the nearest neighboring cells. 
This does not necessarily imply that particles diffuse the distance 
of a coarse grain cell. It is just the most efficient way 
to simulate mass transfer without considering all the details 
of the diffusion process.
This diffusion process moderates the height fluctuation that would arise 
from a completely random deposition process, and enables the surface to grow 
in a layer-by-layer growth mode. Such local diffusion process was 
first introduced in a model proposed by Family \cite{family}, 
which results in the reduction (smoothing) of the surface height 
fluctuations, $i.e.$, 
the surface of the material simulated in this way exhibits the 
layer-by-layer growth up to some characteristic thickness, after 
which the height fluctuation begins to increase with time\cite{damping}. 
Our simulations are thus limited to a characteristic thickness, 
restricted by the above layer-by-layer growth requirement. 
Although the above model of the eventual emergence of surface 
roughness is universal, being independent 
of the diffusion length, in our simulation we restrict the diffusion 
length to the lattice constant of the unit cell. 
It is worth noting that the MC simulations using coarse-grain 
particles lead to the same result as that obtained with a 
single-particle model in the problem of the island density 
distribution in monolayer epitaxial growth\cite{wolf}.\\ 

$Simulation$ $results$: A typical superlattice generated by 
the MC simulations just described is shown in Fig.2a. 
This representation does not by itself show a clear separation 
between the predominantly ZnSe- and ZnTe-rich regions. 
In order to increase the contrast in the modulation 
pattern, in Fig.2b we replot the same data, but with the alloy 
density averaged along the $y$ direction in each $(x,z)$ point. 
The modulation pattern can now be seen distinctly. 
To bring out the modulation more clearly, we next consider the 
density profile of one compound (ZnSe), as a function of $z$, 
measuring the concentration of ZnSe within a square of 
$\ell_x \times \ell_y = 10a \times 10a$, 
located at the edge of the substrate. \\ 

As shown in Figs.3a and 3b, the concentration exhibits two 
different types of sinusoidally-oscillating 
behavior. When the angular velocity $\omega$ of the rotation 
of the substrate is small, the amplitude of the modulation pattern 
is invariant for different heights $z$ (Fig.3a). On the other hand, 
for large $\omega$, the envelope of the modulation pattern is 
itself modulated (Fig.3b). The latter behavior 
occurs when the period of the rotation of the substrate  
$T_r=2\pi/\omega$ is short, and is not integer-multiple of 
the time needed to deposit one monolayer $T_m=1/(f_1+f_2)$. 
While the epilayer grows by one monolayer, the substrate rotates with 
the angle $\theta_m=\omega T_m$. When $\theta_m \ne 2\pi/n$ for integer 
$n$, the second type of the density profile occurs; 
when $\omega$ is not large enough, the top layer is only partially 
covered during the period of one revolution. 
The fraction of covered cells changes as the revolution proceeds 
further. When $\theta_m \ne 2\pi/n$, the fraction of covered cells 
on the top layer after $n$ revolutions of the substrate is not the same 
as the one before. Let $n$ be the number of monolayers between 
two successive peaks in the density profile as seen in Fig.3b. Then since 
the fractions of covered cells before and after the $n$ revolutions 
are not equal, the densities of one compound at the positions 
of the two peaks are different, too.\\
 
In particular, when the ratio between the two time scales 
$T_r/T_m=(2\pi/\omega)(f_1+f_2)$ can be written as a 
rational fraction $k/\ell$, where $k$ and $\ell$ are integers, 
the substrate returns to the same orientation with the same 
fraction of covered cells after $\ell$ revolutions of the substrate, 
during which $k$ monolayers are deposited. 
Thus, the amplitude of the envelope is recovered to its 
value after $\ell$ revolutions of the substrate. 
For example, when $\omega=3.0$ and $f_1+f_2 \approx 1.854$, 
the ratio becomes $T_r/T_m\approx 31/8$. Therefore the period 
of envelop modulation is 8 revolutions, during which 31 
monolayers of material is deposited, as shown in Fig.3b.\\

Next, we investigated the power spectrum of the density profile, 
corresponding to the x-ray scattering intensity observed 
experimentally. 
We found that the power spectrum shows a single peak located at 
$\omega_p$ as shown in Fig.4, regardless of the two types of 
the density profile. The location of the single peak is 
proportional to the angular velocity of the substrate rotation, 
as shown in Fig.5. The occurrence of the single peak in the 
power spectrum, even for the case when the modulated envelope is 
present, implies that there exists only one generic frequency 
in the system. This generic frequency is equivalent to the angular 
frequency of the rotation of the substrate. To understand 
why only a single peak occurs even for the case with the 
modulated amplitude, we in Fig.6 replot the data of in Fig.3a using  
three-monolayer unit rather than one-monolayer unit 
used in Fig.3a. The amplitude of the modulation is itself 
modulated, even though Fig.3a and Fig.6 are drawn from 
identical data. Since the two density profiles are actually 
identical, their power spectra would be the same, exhibiting 
one single peak, located at the position corresponding to the 
angular frequency $\omega$, which is $\omega=1.0$. Next, 
when $x$-axis is rescaled by $1/3$ in Fig.6, the density profile 
becomes equivalent to that in Fig.3b for $\omega=3.0$, but the 
amplitude is reduced. Therefore, it is natural to expect that the 
power spectrum of the density profile in Fig.3b exhibits a single 
peak instead of the two peaks that might intuitively be expected 
to result from the beats of the periods of rotation of the 
substrate and of the amplitude modulation.\\ 
 
Finally, we studied the intensity of the power spectrum, and found that 
it depends on the angular velocity of the substrate $\omega$ and the 
location of the square where the measurement was taken. Here, we 
first examined the intensity as a function of the angular velocity 
$\omega$. As shown in Fig.7, it is found that the modulation 
intensity decreases as the 
angular velocity increases, reflecting that the flux inhomogeneity 
becomes weaker as the angular velocity increases. 
Thus, while we can obtain effective superlattices with 
shorter periods we increase the angular velocity of the rotation of the 
substrate, the flux inhomogeneity incident on a given cell becomes 
weaker, lowering the the quality of the superlattice fabricated 
with higher angular velocity. 
Second, we examined the amplitude of the modulation pattern 
as a function of the location where the measurement was taken. 
As shown in Fig.3a, for a given angular velocity, the amplitude 
gradually decreases to zero as the measurement position approaches 
the center of the substrate. The amplitude has extrema at the 
edges of the substrate, which for our parameters has values 
$\approx 0.62$ and $\approx 0.39$ for the ZnSe content, in 
reasonable agreement with the experimental results.\\ 
 
\vskip -0.4cm
In conclusions, we have performed MC simulations for the 
fabrication of ZnSe$_{1-x}$Te$_x$ SMSL in which the composition 
$x$ that varies sinusoidally. It is found that the modulation of the density 
of ZnSe or ZnTe along the growth direction is caused by both the 
inhomogeneity of the fluxes from the Zn, Se, 
and Te sources incident at a given point of the substrate and by 
the rotation of that substrate. 
The modulation of the structure is displayed in Figs.2 and 3, 
showing the alternative occurrence of gradually-changing ZnSe-rich 
and ZnTe-rich domains along the growth direction, instead of sharp 
interfaces between these domains. When the rate of substrate rotation is slow 
(fast), the amplitude of the modulation pattern is monotonic 
(modulated). It is found that the estimated period of the 
modulation pattern is proportional to the rate of rotation. 
The mean modulation amplitude decreases as the rate 
of rotation of the substrate increases, and also as the 
measurement position approaches the center of the substrate.\\
 
\vskip -0.5cm
This work was supported by the U.S. DOE Grant 97ER45644 and 
the Korean Research Foundation (Grant No. 99-041-D00150). 


\newpage 

{\bf FIGURE CAPTIONS}

\begin{itemize}

\item Figure 1. Schematic view of the MBE chamber, showing the 
relative positions of the sources and the size of the substrate.

\item Figure 2. ``Snapshot" of the SMSL structure for $\omega=0.5$ rad/sec 
with (a) raw MC data, and (b) processed data using the averaged 
density along the $y$ direction, as discussed in the text.

\item Figure 3. The profile of the density of the ZnSe compound as 
a function of height $z$ for $\omega=0.5$ (a) and 3.0(b). 
In (a), the data with the largest amplitude of modulation (solid line), 
median (dotted line), and smallest amplitudes (dashed line) are 
selected from the squares of $10a\times 10a$ ($a=0.2$mm), whose 
centers are located at $(x,y)=(-135a,0)$, 
$(-75a,0)$, and $(-45a,0)$, respectively. 
The symbol ($\diamond$) denotes the densities obtained at one-monolayer 
steps $T_m$. The data are averaged over 100 configurations.  

\item Figure 4. The power spectrum of the density profile of the 
ZnSe compound for $\omega=1.0$ and 3.0 (inset). 
We perform a Fourier transform of the density profile with 
1024 data points, resulting in the frequency ranged 
from 1 to 512.  

\item Figure 5. Plot of the estimated location of the Fourier peak in the power 
spectrum versus the rate of substrate rotation, $\omega$, showing 
that only the $\omega$-parameter determines the power-spectrum 
distribution.  

\item Figure 6. A plot of the density profile for Fig.3a using 
three-monolayer steps ($\diamond$).  

\item Figure 7. Modulation amplitude of SMSL (in terms of ZnSe density) 
as a function of the rate of substrate rotation $\omega$. 
For highest values of $\omega$ the short-period SMSL gradually 
transforms into a homogeneous alloy.  
\end{itemize}

\centerline{\epsfxsize=7.3cm \epsfbox{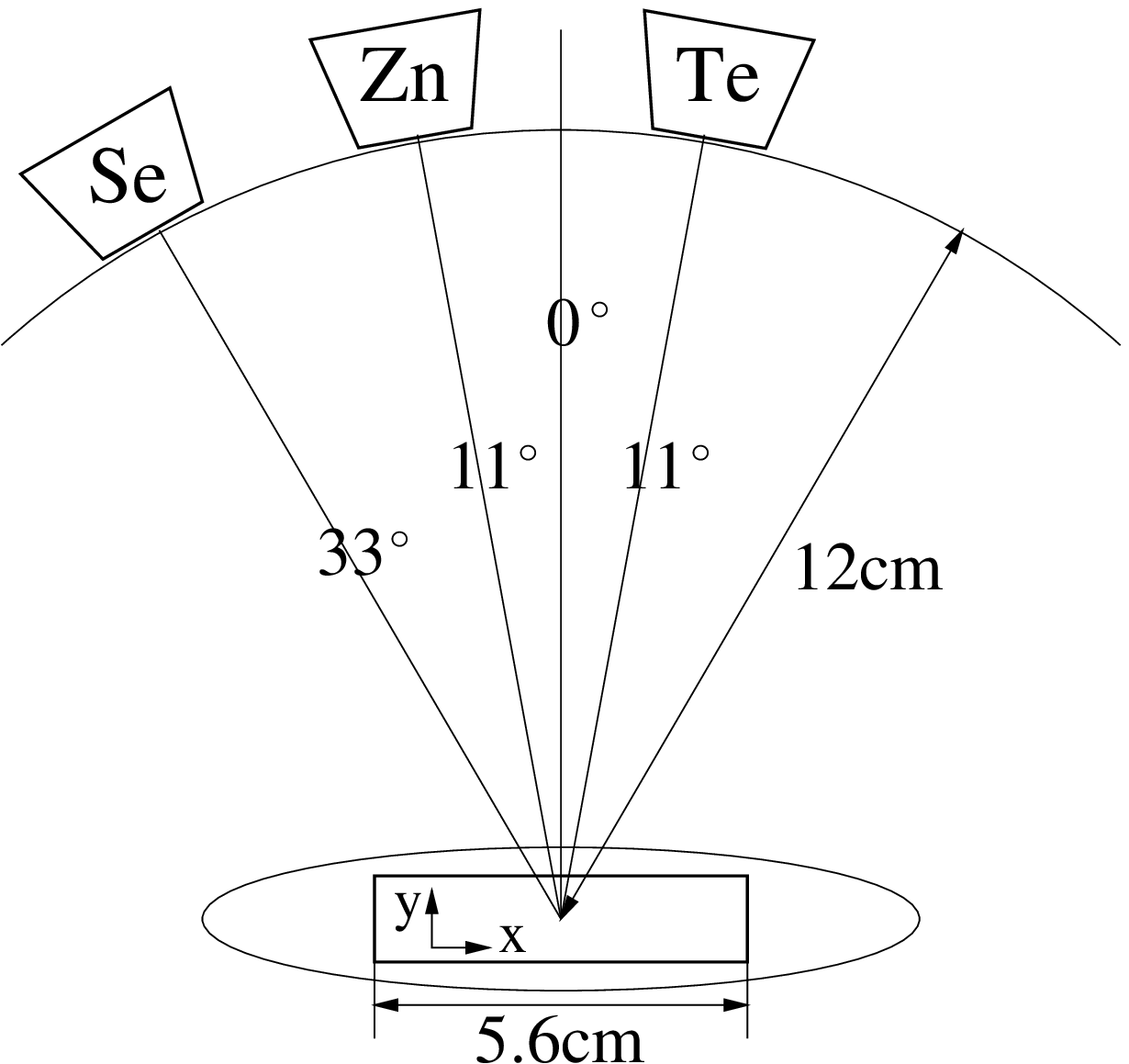}}
\vfill
Figure 1. H. Jeong {\it et al.} ``Monte Carlo Simulation of Sinusoidally Modulated Superlattice Growth'' 

\newpage

\centerline{\epsfxsize=9.3cm \epsfbox{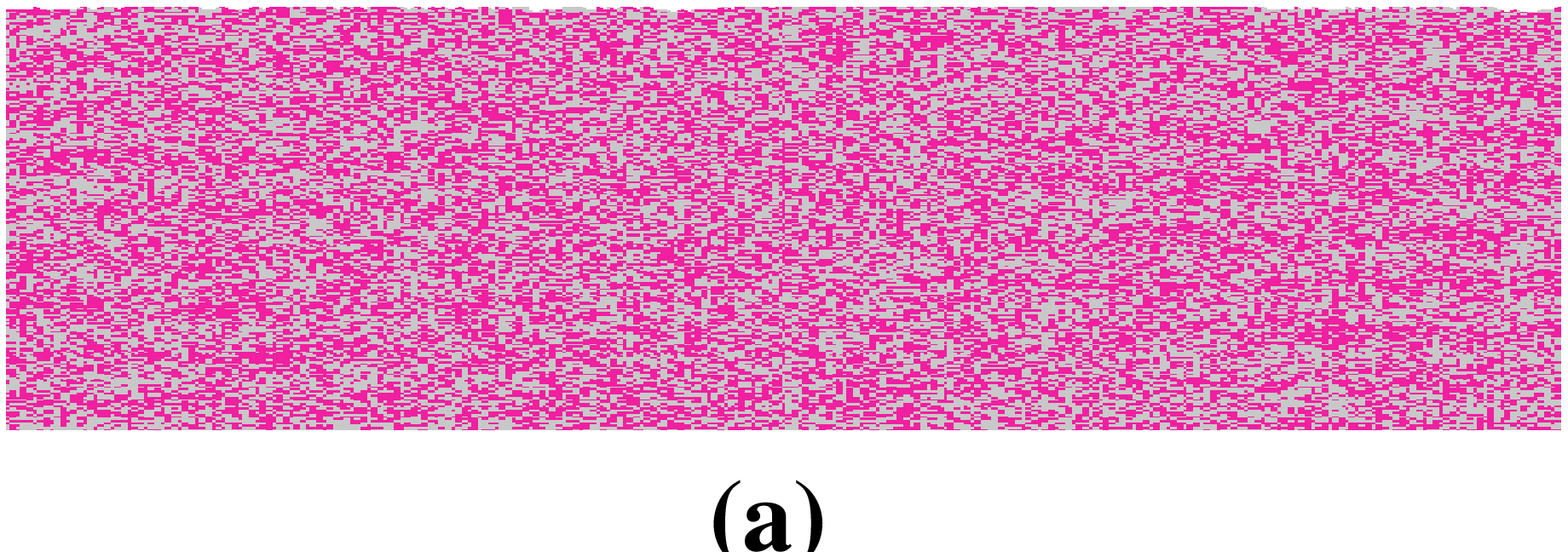}}
\centerline{\epsfxsize=9.3cm \epsfbox{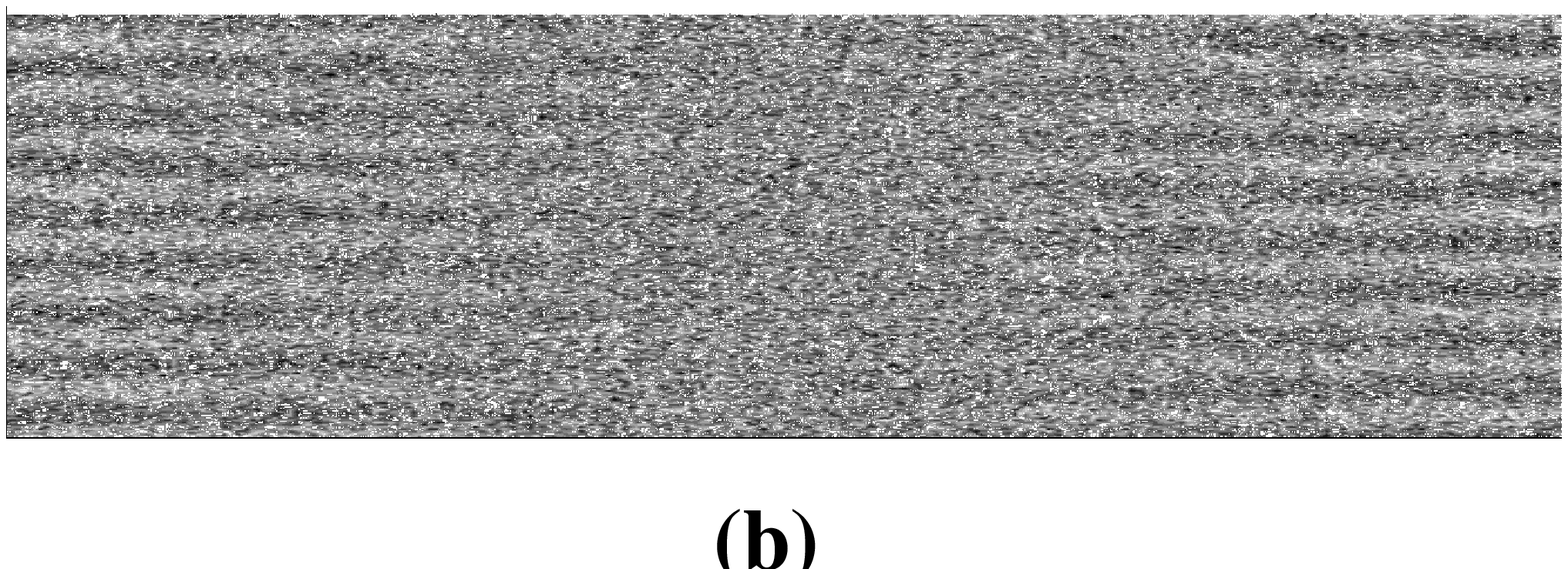}}
\vfill
Figure 2. H. Jeong {\it et al.} ``Monte Carlo Simulation of Sinusoidally Modulated Superlattice Growth'' 

\newpage

\centerline{\epsfxsize=9.3cm \epsfbox{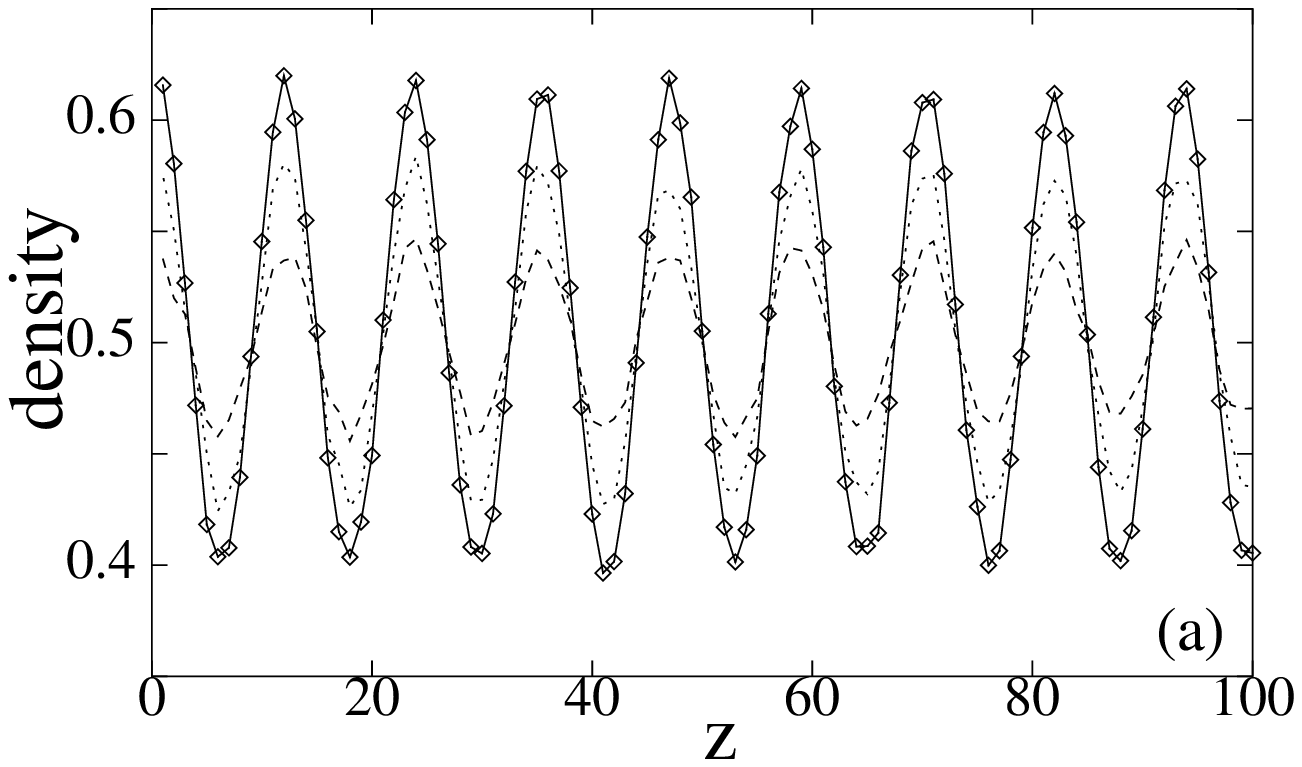}}
\centerline{\epsfxsize=9.3cm \epsfbox{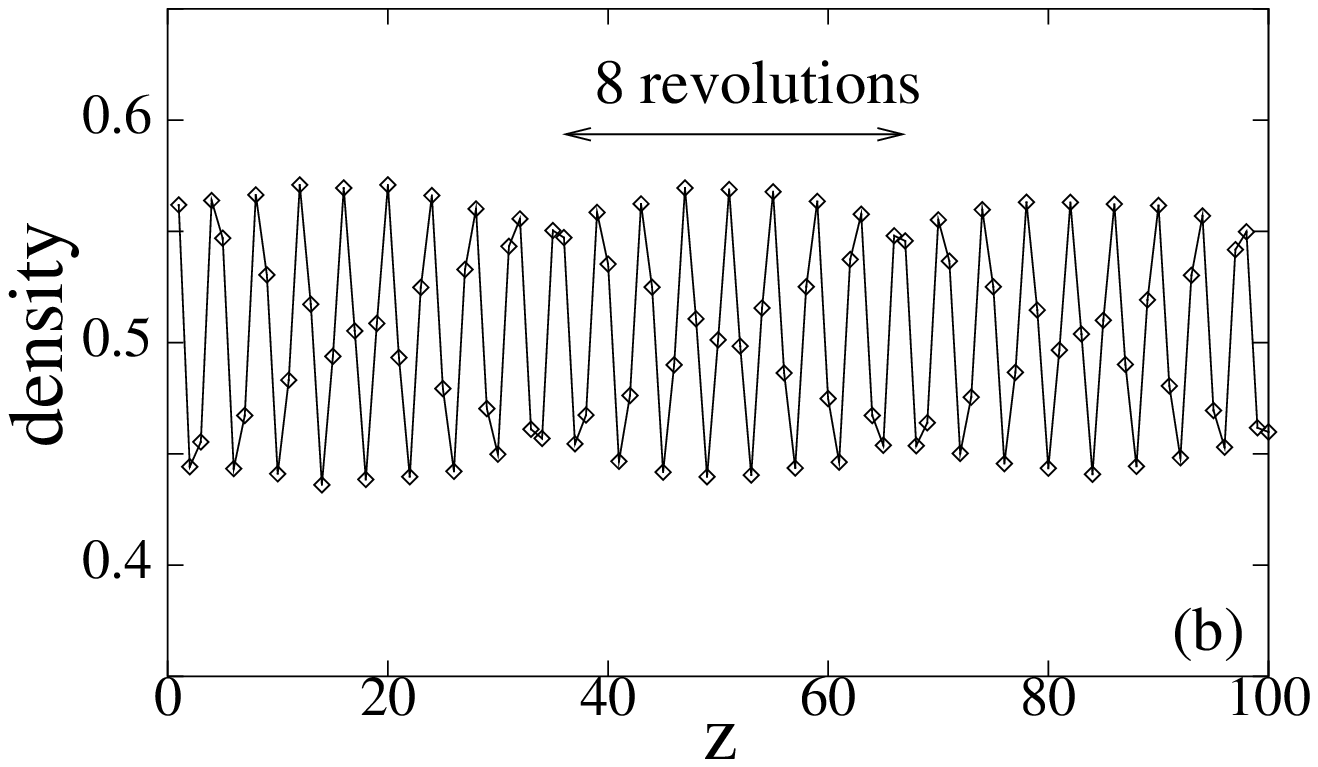}}
\vfill
Figure 3. H. Jeong {\it et al.} ``Monte Carlo Simulation of Sinusoidally Modulated Superlattice Growth'' 

\newpage

\centerline{\epsfxsize=9.3cm \epsfbox{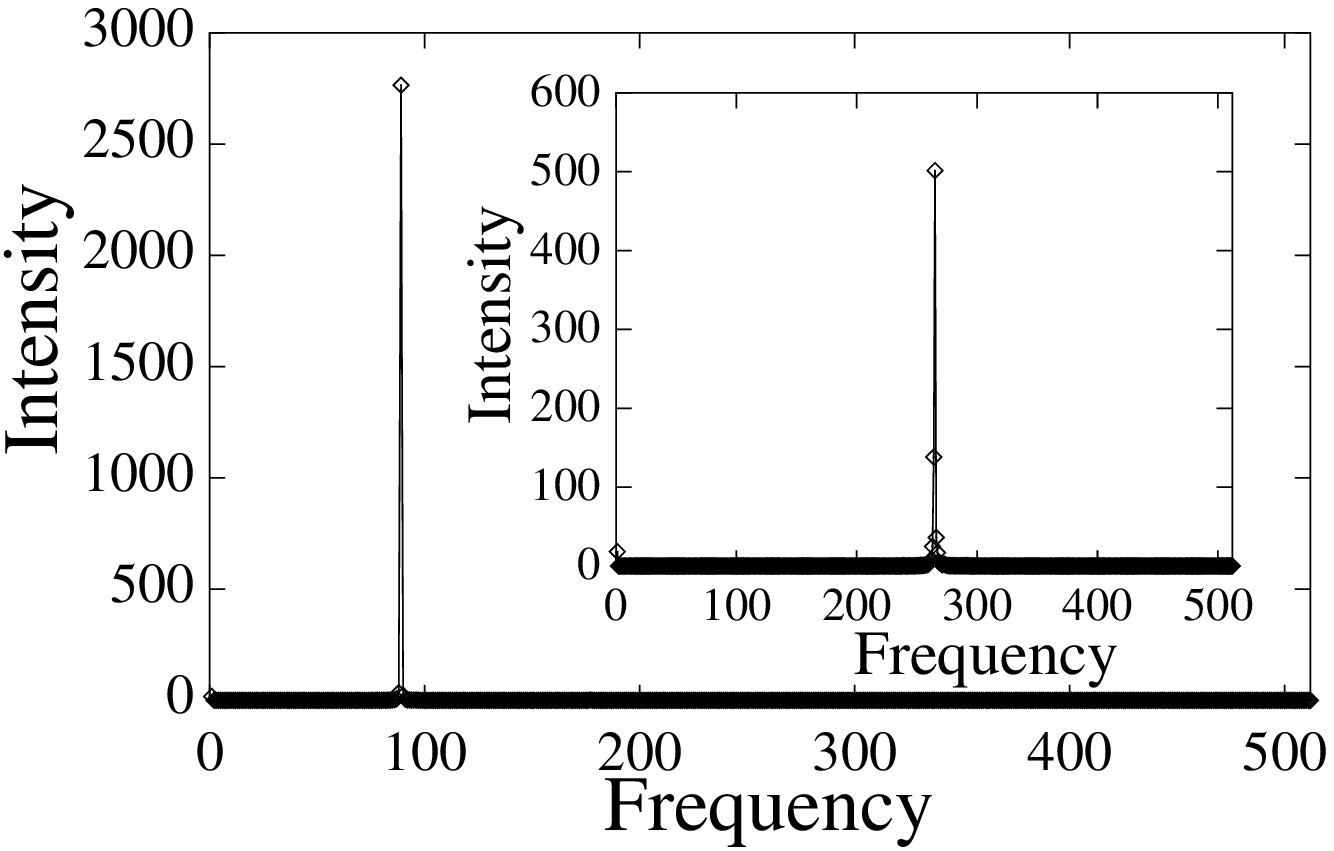}}
\vfill
Figure 4. H. Jeong {\it et al.} ``Monte Carlo Simulation of Sinusoidally Modulated Superlattice Growth'' 

\newpage

\centerline{\epsfxsize=9.3cm \epsfbox{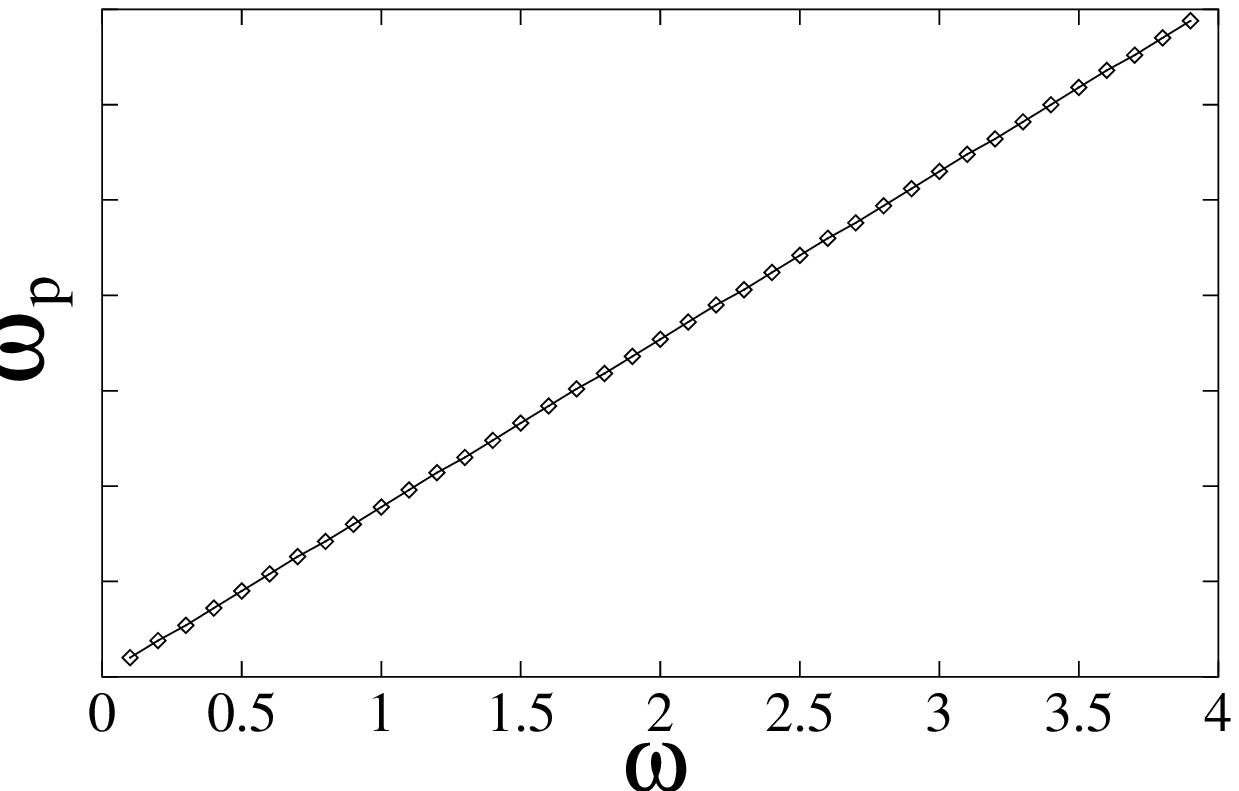}}
\vfill
Figure 5. H. Jeong {\it et al.} ``Monte Carlo Simulation of Sinusoidally Modulated Superlattice Growth'' 

\newpage

\centerline{\epsfxsize=9.3cm \epsfbox{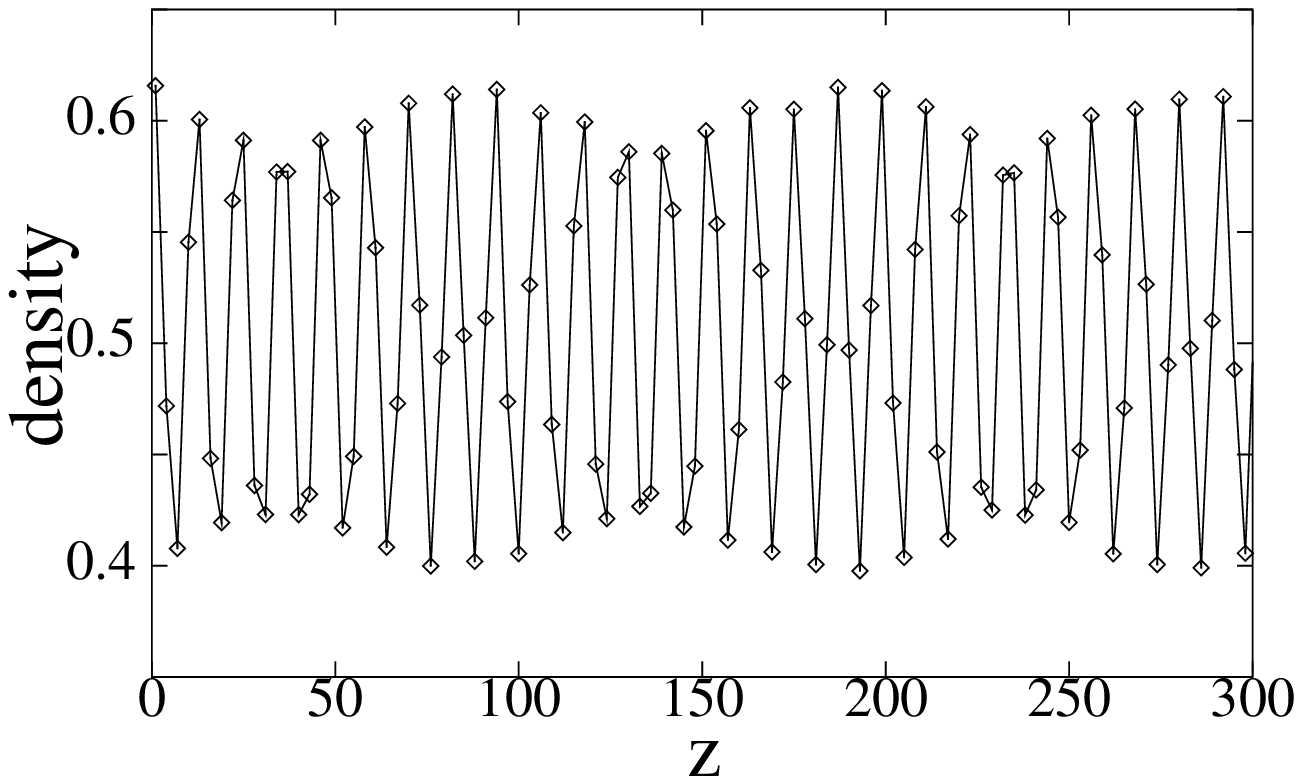}}
\vfill
Figure 6. H. Jeong {\it et al.} ``Monte Carlo Simulation of Sinusoidally Modulated Superlattice Growth'' 

\newpage

\centerline{\epsfxsize=9.3cm \epsfbox{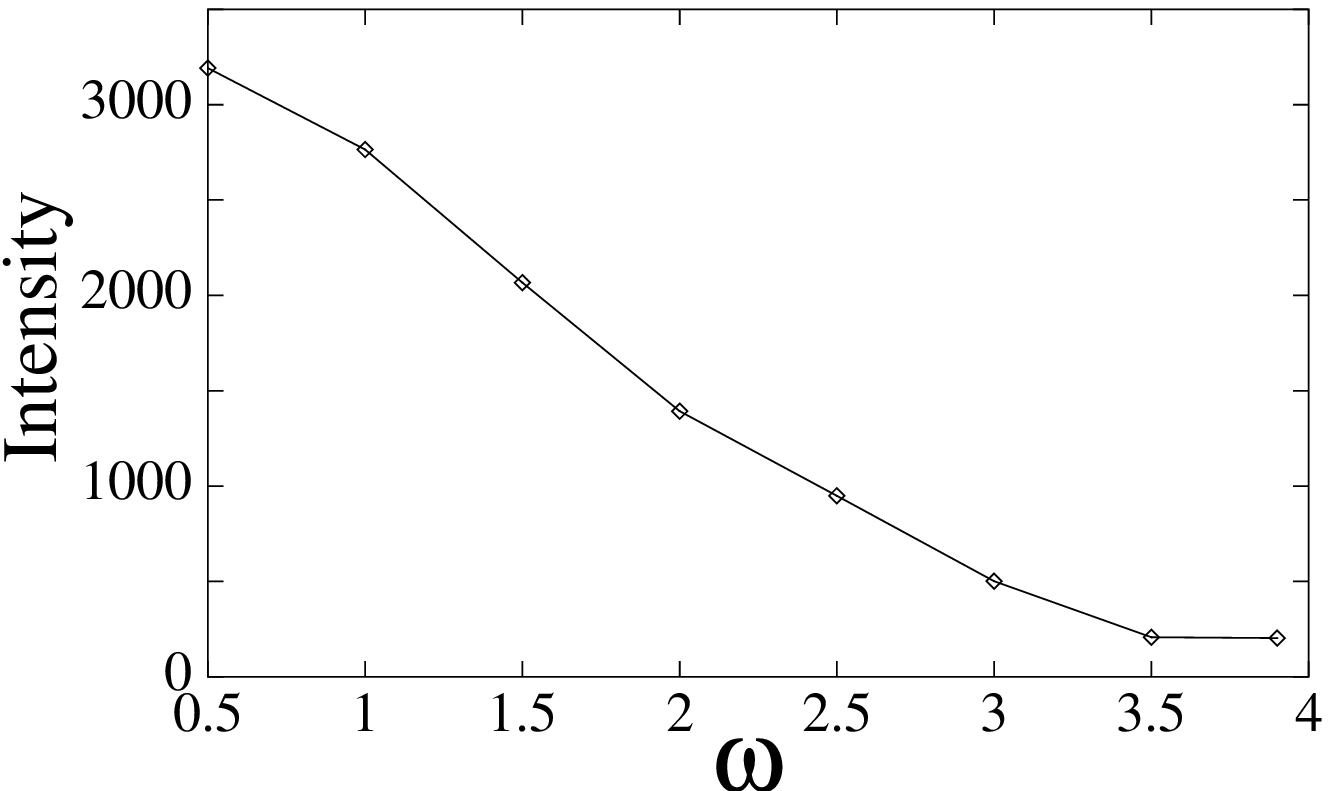}}
\vfill
Figure 7. H. Jeong {\it et al.} ``Monte Carlo Simulation of Sinusoidally Modulated Superlattice Growth''


\begin{thebibliography}{99}
\bibitem[*]{corres} Author to whom correspondence should be 
addressed. Email: kahng@phya.snu.ac.kr
\bibitem{prb1} P.M. Reimer, J. R. Buschert, 
S. Lee, and J.K. Furdyna, Phys. Rev. B {\bf 61}, 8388 (2000). 
\bibitem{prb2} S. Lee, U. Bindley, J.K. Furdyna, P.M. Reimer, 
and J.R. Buschert, J. Vac. Sci. Technol. B {\bf 18}, 1518 (2000). 
\bibitem{mgs} K.T. Obinata, I. Suemune, H. Kumano, 
and J. Nakahara, {\rm Appl. Phys. Lett.} {\bf 68}, 844 (1996).
\bibitem{znsse} K. Shahzad, D.J. Olego, and C.G. Van de Walle, 
{\rm Phys. Rev. B} {\bf 38,} 1417 (1988).
\bibitem{xray} P.F. Fewster, {\rm Semicond. Sci. Technol.} 
{\bf 8} 1915 (1993). 
\bibitem{prb3} G. Yang, S. Lee, and J.K. Furdyna, Phys. Rev. B {\bf 61}, 
10,978 (2000). 
\bibitem{wolf} D.E. Wolf, in {\it Scale Invariance, Interfaces, 
and Non-Equilibrium Dynamics,} edited by A. McKane {\it et al.} 
(Plenum, New York, 1995), pp. 215-248. 
\bibitem{family} F. Family, {\rm J. Phys. A} {\bf 19}, L441 (1986).
\bibitem{damping} S. Park, H. Jeong, and B. Kahng, {\rm Phys. 
Rev. E} {\bf 59,} 6184 (1999). 
\end{thebibliography}
\end{document}